# Experimental test of magnetic photons

R. S. Lakes, Ph. D.

Department of Engineering Physics
Engineering Mechanics Program; Biomedical Engineering Department
Materials Science Program and Rheology Research Center
University of Wisconsin-Madison
147 Engineering Research Building
1500 Engineering Drive, Madison, WI 53706-1609
http://silver.neep.wisc.edu/~lakes

28 May 2004

**Abstract**
A "magnetic" photon hypothesis associated with magnetic monopoles is tested experimentally. These photons are predicted to easily penetrate metal. Experimentally the optical transmittance T of a metal foil was less than $2 \times 10^{-17}$. The hypothesis is not supported since it predicts $T = 2 \times 10^{-12}$.

The intriguing possibility of a second, penetrating form of light is allowed within several variants of classical electromagnetism, in particular nonzero photon mass or the existence of a second "magnetic" photon associated with the theory of monopoles.

Electromagnetism in the presence of nonzero photon mass is described by the Maxwell-Proca equations [1,2] in free space.

$$\text{div } \mathbf{E} = 4\pi\rho - \mu^2 V \qquad \text{curl } \mathbf{E} = -\frac{1}{c}\frac{\partial \mathbf{B}}{\partial t} \qquad (1a,b)$$

$$\text{div } \mathbf{B} = 0 \qquad \text{curl } \mathbf{B} = \frac{1}{c}\frac{\partial \mathbf{E}}{\partial t} + \frac{4\pi}{c}\mathbf{J} - \mu^2 \mathbf{A} \qquad (2a,b)$$

in which **E** is electric field, **B** is magnetic field, $\rho$ is charge density, **J** is current density, V is the scalar potential, **A** is the vector potential, c is the speed of light, and $\mu^{-1} = \hbar/m c$ is a characteristic length, the Compton wavelength of the photon, with m as the photon mass. A nonzero photon mass would, among other effects, give rise to a wavelength dependence of the speed of light in free space, and the possibility of longitudinal electromagnetic waves. Such longitudinal waves are predicted to be created when a transverse wave is obliquely incident on an interface between dissimilar materials [3], and they are partially converted back into transverse waves by interaction with another interface. The conversion efficiency is small for any nonzero photon mass consistent with experiment. Since the characteristic length is so large, photon mass is difficult to measure; either one conducts a laboratory scale experiment with extremely high precision, or one draws inference from astrophysical observation on a large scale. Since the wavelength of visible light is so short compared with any reasonable Compton wavelength for the photon, attempts to generate and detect longitudinal light must be far less sensitive than quasistatic methods. Recently a laboratory experiment [4] was done to detect effects of cosmic magnetic vector potential on a toroid. That and a more recent experiment [5] of similar design set limits on photon mass but did not detect any effects.

If magnetic monopoles exist, Maxwell's equations assume a symmetric form [6] which includes both electric $\rho_e$ and magnetic $\rho_m$ charge density and corresponding current densities $\mathbf{J}_e, \mathbf{J}_m$.

$$\text{div } \mathbf{E} = 4\pi\rho_e \qquad \text{curl } \mathbf{E} = -\frac{1}{c}\frac{\partial \mathbf{B}}{\partial t} - \frac{4\pi}{c}\mathbf{J}_m \qquad (3a,b)$$

$$\text{div } \mathbf{B} = 4\pi\rho_m \qquad \text{curl } \mathbf{B} = \frac{1}{c}\frac{\partial \mathbf{E}}{\partial t} + \frac{4\pi}{c}\mathbf{J}_e \qquad (4a,b)$$

Although no monopoles have been observed, some monopole theories predict observable interactions between magnetic photons and matter without any monopoles near the observer. Salam (1966) [7] considered two variants of monopole theory, one which involved large unphysical C-violating effects in atomic physics, and the other involving two kinds of photons, electric and magnetic. The electric photons interact with leptons and hadrons, and the magnetic photons would interact with hadrons but not leptons. This magnetic photon concept was critiqued [8,9] in view of the fact the electric and magnetic sources become independent, so one loses the Dirac [10] quantization condition. Monopole theories without a second magnetic photon give a static pole strength of zero [8] or are noncovariant [9]. Moreover, the Dirac monopole and finite photon mass cannot coexist within the same theory [11]. More recently, a model of magnetic monopoles was presented [12] in which the magnetic photon couples tensorially, interpreted as coupling with the velocity, of electric charges with respect to the cosmic background. This theory predicts that creation, shielding, and absorption of magnetic photons is suppressed by a factor of $7 \times 10^5$ compared with normal electric photons of the same energy. The magnetic photons are predicted to penetrate metal since the penetration depth for visible light is a few nanometers, so the corresponding depth for magnetic photons is expected to be a few millimeters. Since both creation and detection of magnetic photons are suppressed by the same factor, the predicted intensity of light detected after passage through a metal foil is a factor $2 \times 10^{-12}$ weaker than the incident beam. The present work is directed toward experimental evaluation of such effects.

A preliminary experiment was conducted using an amplified silicon detector (Newport, type 834 optical power meter). A 600 W halogen projector bulb was set up with a fan for cooling. A 125 mm diameter lens system derived from a TV projection system was used to focus the light to a spot about 1 cm in size. The power inferred over the 1 cm$^2$ detector area was about 13 W. This detector is sensitive to both visible light and near infrared. The detector was prepared without the usual attenuator filter; the face was covered with a single layer of aluminum foil of thickness 13 μm. Additional layers were taped around the sides to exclude stray light. The display was observed as the lamp was turned on and off. No effect was seen, within the instrument's resolution of 100 fW. The foil's transmittance limit is therefore $T < 8 \times 10^{-15}$, a value lower than predicted by a factor of at least 250.

Experiments were conducted using an MH983 channel type photomultiplier (Perkin-Elmer Optoelectronics, Fremont, CA) to detect light. The claimed quantum efficiency is about 15% at a wavelength of 450 nm, 7.5% at 532 nm and 2% at 600 nm. The sensitive photocathode is about 5 mm in diameter. The photomultiplier face was covered with a single layer of aluminum foil of thickness 13 μm. The foil was secured with black tape and additional layers were added to the sides of the device to exclude stray light. The photomultiplier was then clamped to a support upon an isolated optical table. A beam from an 80 mW diode-pumped YAG laser at 532 nm was directed at the center of the foil covered recessed photomultiplier face. The pulse output of the photomultiplier was input into an oscilloscope and a digital counter (Fluke 1953A). Data were collected in groups of six trials of ten seconds each, with the laser off and with the laser on. The average count rate with the laser off was 3.2 ± 0.25 counts per second. With the laser on, two such runs yielded 4.4 ± 0.95 and 3.8 ± 0.45. A subsequent set of six trials with the laser off yielded 3.7 ± 0.27 counts per second. Here ± refers to the standard deviation. These count rates are comparable to the manufacturer's claimed dark count rates of about 3 per second "typical" (at 20 °C) and 5 for the present unit (at 22 °C). A transmittance of $2 \times 10^{-12}$ based on [12] of 80 mW green light detected with a 7.5% quantum efficiency would give $3.2 \times 10^4$ counts per second. The results are consistent with between 0 and 0.3 counts per second over noise. The transmittance limit is therefore $T < 2 \times 10^{-17}$, a value lower than predicted by five orders of magnitude.

A further test was conducted using a 50 W halogen projector lamp (with reflector) as a source and the photomultiplier as a detector. The rationale is to allow for the possibility that sources may not be equivalent in the hypothetical emission of magnetic photons. The visible intensity (measured with a filtered silicon sensor) was about 160 mW/cm$^2$ at 70 mm from the lamp, corresponding to about 30 mW over the 5 mm diameter sensitive cathode. This is a somewhat less stringent test than the one with the laser in view of the lower power and the fact the photomultiplier has limited sensitivity in the red region dominated by the halogen lamp. Six trials of ten seconds with the lamp on yielded 2.7 ± 0.67 counts per second. This rate is comparable to the dark count rate measured above, with no laser light. The lamp was turned off and six further measurements of ten seconds each yielded 5.1 ± 0.63 counts per second. The slight increase in count rate with the lamp off, though still within the range of accepted dark count rate for the device, was attributed to the *known* tendency of photomultipliers to increase their dark count with temperature, since the optical and infrared output of the lamp was sufficient to warm the front of the sensor. Indeed, that is why photomultipliers are cooled for high performance applications. A temperature effect on the dark count is also the likely cause for the slight increase in dark count after the above laser experiments in comparison with the dark count before.

In summary, the experimental results do not support the theory suggested in [12] that magnetic photons which couple with charges via a velocity term give rise to a second form of light. That theory has no adjustable parameters. The experiments are more sensitive by a factor of $10^5$ than needed to detect such magnetic photons.

There are several monopole theories involving magnetic photons. In addition to those of Ref. [7], [12], supersymmetric SU(N) gauge theories can predict gluino condensate in which the superpotential gives a mass to the dual (magnetic) photon [13]. A U(1) theory including magnetic photons was developed to predict monopole pair production [14]. Chiral components of a rank-2 spinor field were taken as the dynamic variables of a monopole theory which accommodates the magnetic photon [15] which interacts only with monopole charges and currents. Implications of most such theories in the absence of monopoles were not explored in detail. Therefore we have not excluded the possibility of the existence or detectability of magnetic photons within a theoretical framework different from that in [12].